\documentstyle[sprocl]{article}
\def\Journal#1#2#3#4{{#1} {\bf #2}, #3 (#4)}

\def\PRD{{\em Phys. Rev.} D}
\def\PLB{{\em Phys. Lett.} B}
\def\ZPA{{\em Z. Phys.} A}
\def\PR{\em Phys. Rev.}
\def\IJMP{{\em Int. J. Mod. Phys.} A}
\def\Er{(E) A}

\def\vep{\varepsilon}
\def\ko{K^0}
\def\kb{\overline{K^0}}
\def\bo{B^0}
\def\bb{\overline{B^0}}
\def\be{\begin{equation}}
\def\ee{\end{equation}}

\begin{document}

\title{
$CP-$VIOLATING PARAMETERS FOR NEUTRAL $B-$MESONS
AND THEIR COMPLETE MEASUREMENT
}

\author{Ya. AZIMOV}

\address{Petersburg Nuclear Physics Institute\\
Gatchina, St.Petersburg, 188350, Russia\\
e-mail: azimov@pa1400.spb.edu}

\maketitle
\abstracts{
Phenomenological $CP$-violating parameters in decays of neutral
$B$-mesons are discussed with special attention to the degree of
their measurability. Important role of the sign of $\Delta m_B$
is emphasized. We briefly describe how it could be determined
experimentally.  }

\section {Introduction}
Twenty years ago the statement~\cite{ans}:
\begin{itemize}
\item The origin of $CP$-violation will not be established till
its manifestations are known only for neutral kaons
\end{itemize}
could be considered as somewhat heretical. Now it and one more
statement:
\begin{itemize}
\item  The most promising testing ground for detailed studies of
$CP$-violation is provided by decays of neutral $B$-mesons,
\end{itemize}
became generally accepted (see, e.g., reviews~\cite{bk,nq,br}).

As a result, many papers have been devoted to discussion of
$B$-meson decay modes favorable for $CP$-violation searches and
to experimental manifestations of possible sources of the
violation (see, e.g., references in reviews~\cite{bk,nq,br}).
A more straightforward problem, degree of measurability of
phenomenological parameters describing $CP$-violation in
$B$-meson decays, has not been considered. One possible reason
could be a close similarity of neutral $B$-mesons to neutral
kaons. However, heavier masses of the third quark generation
produce various differences, sometimes rather essential, in the
meson decay properties. Here we discuss basic $CP$-violating
parameters for neutral $B$-mesons with special attention to the
question how one could achieve their complete measurement. The
presentation is essentially based on the papers~\cite{a,ars}.

\section{Standard $CP-$violating parameters}

Time evolution of neutral $B$-mesons is known to be
determined by the two eigenstates
\be
B_\pm\ =\ \frac{1}{\sqrt{2(1+|\vep_B|^2)}}\
\left[(1+\vep_B)\bo\pm(1-\vep_B)\bb\right]\ .
\ee
If we apply the phase convention $\bb=(CP)\bo$, exact
$CP$-conservation would imply $\vep_B=0$ and the states
$B_\pm$ having the definite $CP$-parities equal $\pm 1$.
Generally, they are eigenstates of an effective
(non-Hermitian) Hamiltonian. Since $\vep_B$ changes under
rephasing $\bo$ and $\bb$ (without influencing $B_\pm$), it
cannot be measurable itself. Only $\left |\frac{1+\vep_B}{1-
\vep_B}\right |$ is rephasing-invariant and admits measurement
in experiment. The value
\be \delta_B\ =\ \frac{|1+\vep_B|^2-|1-\vep_B|^2}{|1+\vep_B|^2+
|1-\vep_B|^2}\ =\ \frac{2{\rm Re}\,\vep_B}{1+|\vep_B|^2}\ ,
\ee
directly similar to the quantity $\delta_K$ for neutral kaons,
may be considered as the measure of $CP$-violation in
$\bo\bb$ mixing.

The Standard Model leads to an extremely small, really
unmeasurable, value of $\delta_B$ (see, e.g., discussion in
Ref.~\cite{auk}), much smaller than $\delta_K$. More promising are
studies of decays
\be
\bo(\bb)\ \rightarrow\ f
\ee
with final states $f$ of definite $CP$-parities~\cite{auk,dr}. To
measure $CP$-violation in a particular decay mode one can use
deviation of the parameter
\be
\lambda_B^{(f)}\ =\ \frac{1-\vep_B}{1+\vep_B}\
\cdot\ \frac{\langle f|\bb\rangle}{\langle f|\bo\rangle}
\ee
from the $CP$-parity value of the state $f$. Any
$\lambda_B^{(f)}$ is rephasing-invariant and, hence, its complete
measurement (i.e., of both the absolute value and phase) should
be possible.

Parameters $\lambda_B$ are similar to analogous parameters
$\lambda_K$ in nonleptonic kaon decays which are equivalent to
more familiar parameters $\eta$. Both sets satisfy many relations
if $CPT$ is conserved (see discussion in Ref.~\cite{ars}).
Unitarity gives one more relation for each set~\cite{bs}. Those
relations, together with the fact that a single partial width of
neutral kaons (for the mode $\ko\to (2\pi)_{I=0}\,$) is more than
3 orders above any other, provide the known structure of kaon
decays with only one parameter of $CP$-violation being
independent and large enough for measurement. An essential
preference of $B$-physics is that having many decays with more
comparable probabilities it can reveal many independent
and measurable $CP$-violating parameters.

Standard calculations for the decay (3) with the initially pure
$\bo$-meson lead to the time distribution
$$W_B^{(f)}(t)\ \sim\ \left|\frac{1+\lambda_B^{(f)}}{2}\right|^2
\exp(-\Gamma_+t)+\left|\frac{1-\lambda_B^{(f)}}{2}\right|^2
\exp(-\Gamma_-t) \ $$
\be
+ \ \exp\left(-\ \frac{\Gamma_+ +\Gamma_-}{2}\ t\right)
\left(\frac{1-|\lambda_B^{(f)}|^2} {2}\ \cos\Delta m_Bt-{\rm
Im}\lambda_B^{(f)}\ \sin\Delta m_Bt\right)\
\ee
which exhibits degree of measurability of parameters $\lambda_B$.
Here we denote $\Delta m_B=m_--m_+$; $m_+, \Gamma_+$ and $m_-,
\Gamma_-$ are the mass and width of the corresponding state
$B_\pm$. Eq.(5) has the same structure as, e.g., decay yield of
$\ko(t)\to\pi\pi$.  The first two terms are contributions of the
eigenstates $B_{\pm}$, the last two terms describe their
interference.

Distribution (5) contains contributions of
$|\lambda_B^{(f)}|^2$, ${\rm Re}\lambda_B^{(f)}$ and
${\rm Im}\lambda_B^{(f)}$ multiplied by different functions
of time. So, at first sight, all the three quantities can be
easily extracted if the distribution is found experimentally
with necessary accuracy. It is just the case in two-pion decays
of neutral kaons where parameters $\lambda_K^{(\pi\pi)}$ have
been completely measured indeed.

But ${\rm Re}\lambda_B^{(f)}$ would not appear at all in
distribution (5) if $\Gamma_+$ and $\Gamma_-$ coincided. So, a
very small expected difference of $\Gamma_+$ and $\Gamma_-$,
contrary to neutral kaons, may prevent direct measurement of
${\rm Re}\lambda_B^{(f)}$. Indirect measurement is still
possible, of course, since $|{\rm Re}\lambda_B^{(f)}|$ can be
calculated from $|\lambda_B^{(f)}|$ and $|{\rm Im}
\lambda_B^{(f)}|$, and the sign of ${\rm Re}\lambda_B^{(f)}$
may be fixed by choosing approximate $CP$-parities of the neutral
$B$ eigenstates (detailed discussion of the procedure and its
relation to experiments see in Refs.~\cite{a,ars}).

The situation for ${\rm Im}\lambda_B^{(f)}$ is less simple.
Eq.(5) contains it multiplied by $\sin\Delta m_Bt$, which sign is
still unknown since only $|\Delta m_B|$ has been measured. For
kaons, special experiments on $K_S$ regeneration in several
plates have allowed to measure the sign of $\Delta m_K=m_L-m_S$
in respect to the known sign of the regeneration phase
(collection of results see in Ref.~\cite{ka}). Similar
experiments for $B$-mesons are impossible because of too small
lifetime. However, without knowing the sign of $\Delta m_B$ one
cannot achieve the complete measurement of $CP$-violating
parameters in neutral $B$-meson decays. Below we discuss how to find
the sign on the base of the method suggested in Ref.~\cite{a}.

\section{Unusual properties of heavy meson decays}

Specific feature of neutral $B$-mesons, having no analogues for
neutral kaons, is the existence of decays
\be
\bo(\bb)\ \to\ f\ko(\kb)\ ,
\ee
with $f$, again, being definite $CP$-parity states. They are
mainly induced by the quark decay $b\to c\overline c s$.
The most popular final state of such a kind is
$J/\psi\ko(\kb)$. Decays (6) generate a new set of
$CP$-violating parameters
\be
\lambda_{BK}^{(f)}\ =\ \frac{1-\vep_B}{1+\vep_B}\
\cdot\ \frac{1+\vep_K}{1-\vep_K}\ \cdot\
\frac{\langle f\kb| \bb\rangle}{\langle
f\ko|\bo\rangle}\ ,
\ee
similar to (4). They are invariant under rephasing of
both $B$ and $K$ mesons and should also be completely measurable.

However, the most interesting and unique property of decays (6)
is the coherence of neutral $B$ and neutral $K$
evolutions~\cite{a}. It leads to double flavor oscillations
which, as was recently emphasized~\cite{ks}, are similar to the
long-known EPR effect~\cite{epr}.

Let us consider, as an example, the cascades
\be
\bo(\bb) \to J/\psi\ko(\kb)\ ,\;\;
\ko(\kb) \to \pi\pi\ (\pi^\mp l^\pm
\nu(\overline\nu) )\ .
\ee
Coherence arises here since only transitions $\bo\to\ko$ and
$\bb\to\kb$ are possible. As a result, kaon evolution is an
immediate continuation of $B$-evolution (though they do not
coincide, of course). This produces unusual properties of such
cascades~\cite{a}. E.g., their double-time distributions over $B$
and $K$ decay times $t_B$ and $t_K$ are non-factorisable. What is
most essential for our present purposes is their sensitivity to
the sign of $\Delta m_B$ in respect to the known sign of $\Delta
m_K$. Corresponding terms in time distributions are generated by
interference at both stages of the evolution (i.e., we need
interference between both $B_+, B_-$ and $K_S, K_L$).

Manifestations of the sign of $\Delta m_B$ have been considered
in more detail for the B-factory~\cite{sar} and LHC~\cite{ars}
environments. Necessary experiments require very high statistics.
Indeed, from available data~\cite{j} we find that any of cascades
(8), appended by decays $J/\psi\to l^+l^-$, has small $({\rm
Br})_{eff}\approx5\cdot 10^{-5}$. Due to necessity of $K_S,K_L$
interference the sign effect always contains an additional small
factor of order $10^{-3}$ ($CP$-violation in two-pion kaon
decays, or small semileptonic branching ratios of $K_S$).
Required statistics seems to be unreachable at the projected
B-factories. LHC may be promising, but statistics of LHC-B at
moderate luminosity~\cite{lhc} also looks insufficient~\cite{ars}.
Effect of the sign of $\Delta m_B$ could be searched for either
by other detectors at LHC or by LHC-B working at full luminosity.
Accurate consideration shows as well that semileptonic kaon decays
in cascades (8) might be more favorable than two-pion ones, but
detailed studies of experimental efficiencies are still necessary.

In summary, we have demonstrated that $CP$-violating parameters
in neutral $B$-meson decays can be completely measured only if
the sign of $\Delta m_B$ is known. This sign might be determined
in special experiments, e.g., at LHC.

\section*{Acknowledgments}
I thank the Organizing Committee of the 2nd International
Conference on B-physics and CP-violation and the Russian
Scientific Program "High Energy Physics" for support of my
participation in the Conference.


\section*{References}
\begin{thebibliography}{99}
\bibitem{ans} A.A. Anselm and Ya.I. Azimov,
\Journal{\PLB}{85}{72}{1979}.
\bibitem{bk} I.I. Bigi, V.A. Khoze, N.G. Uraltsev and A.I. Sanda
in {\em CP Violation}, ed. C. Jarlskog (World Scientific,
Singapore, 1989) p.175.
\bibitem{nq} Y. Nir and H.R. Quinn, {\em Ann. Rev. Nucl. Part.
Sc.}, {\bf42}, 211 (1992).
\bibitem{br} T. Browder, K. Honsheid and D. Pedrini, preprint
UH 511-848-96 (to appear in {\em Ann. Rev. Nucl. Part. Sc.}, vol.
{\bf46}); e-print hep-ph/9606354.
\bibitem{a} Ya.I. Azimov, \Journal{\PRD}{42}{3705}{1990}.
\bibitem{ars} Ya.I. Azimov, V.L. Rappoport and V.V. Sarantsev,
\Journal{\ZPA}{356}{437}{1997}; e-print hep-ph/9608478.
\bibitem{auk} Ya.I. Azimov, N.G. Uraltsev and
V.A. Khoze, {\em Yad. Fiz.} {\bf45}, 1412 (1987) ({\em
Sov. J. Nucl. Phys.} {\bf45}, 878 (1987)).
\bibitem{dr} I. Dunietz and J. Rosner,
\Journal{\PRD}{34}{1404}{1986}.
\bibitem{bs} J.S. Bell and J. Steinberger in {\em Proc. Int. Conf.
on Elementary Particles, Oxford, 1965}, (Chilton, Rutherford High
En. Lab., 1966) p.193.
\bibitem{ka} Particle Data Group, \Journal{\PLB}{170}{132}{1986}.
\bibitem{ks} B. Kayser and L. Stodolsky, preprint MPI-PhT/96-112;
e-print hep-ph/9610522.
\bibitem{epr} A. Einstein, B. Podolsky and N. Rosen,
\Journal{\PR}{47}{777}{1935}.
\bibitem{sar} G.V. Dass and K.V.L. Sarma,
\Journal{\IJMP}{7}{6081}{1992}; \\ \Journal{\Er}{8}{1183}{1993}.
\bibitem{j} Particle Data Group, \Journal{\PRD}{54}{1}{1996}.
\bibitem{lhc} LHC-B, {\em Letter of intent}.
CERN/LHCC 95--5, August 1995.

\end{thebibliography}
\end{document}